\documentclass{aastex62}

\graphicspath{{./}{figures/}}

\received{April 3, 2020}
\revised{July 21, 2020}
\accepted{September 24 2020}

\submitjournal{ApJ}

\shorttitle{How do galaxy properties affect void statistics?}
\shortauthors{Panchal et al.}

\newcommand{\hmpc}{$h^{-1}\mathrm{Mpc}$}

\usepackage{comment}
\usepackage{graphicx}

\begin{document}

\def\illustris{illustris:Nelson-2015,Vogelsberger_2014b,illustris:Vogelsberger-2014,illustris:Genel-2014}

\title{How do galaxy properties affect void statistics?}

\correspondingauthor{Rushy R. Panchal, Alice Pisani}
\email{rpanchal@alumni.princeton.edu, apisani@astro.princeton.edu}

\author{Rushy R. Panchal}
\affiliation{Princeton University, Department of Computer Science,
 35 Olden St, Princeton, 08540, New Jersey, USA}

\author[0000-0002-6146-4437]{Alice Pisani}
\affiliation{Princeton University, Department of Astrophysical Sciences,
4 Ivy Lane, Princeton, 08540, New Jersey, USA}
\author[0000-0002-5151-0006]{David N. Spergel}

\affiliation{Princeton University, Department of Astrophysical Sciences,
4 Ivy Lane, Princeton, 08540, New Jersey, USA}
\affiliation{Center for Computational Astrophysics, Flatiron Institute, 162 5th Avenue, New York, NY 10010, USA}

\begin{abstract}

Using a mapping from dark matter halos to galaxy properties based on hydrodynamical simulations, we explore 
 the impact of galaxy properties on the void size function and the void-galaxy correlation function.
We replicate the properties of galaxies from \texttt{Illustris} on \texttt{MassiveNus} halos, to perform both luminosity and star formation rate cuts on \texttt{MassiveNus} halos. We compare the impact of such cuts on voids properties with respect to cuts on halo mass (as usually performed on halo catalogs driven from N-body simulations). We find that void catalogs built from luminosity-selected galaxies and halos are consistent within errors, while void catalogs built from star formation rate
selected galaxies differ from void catalogs built on halos. We investigate the reason for this difference. 
Our work suggests that voids built on galaxy catalogs (selected through luminosity cut) can be reliably studied by using halos in dark matter simulations.

\end{abstract}


\section{Introduction} 
\label{sec:intro}

Cosmic voids, the large under-dense regions in the galaxy distribution, are a novel probe for cosmology (see \cite{Pisani_2019} and references therein). Due to their low-density nature, they are particularly sensitive probes of dark energy, modified gravity, and neutrinos properties \citep{odrzywolek2009,Lee_2009,Biswas_2010,Lavaux_2010,Lavaux_2012,Bos_2012,Sutter_2012,VIDE, clampitt2013,spolyar2013,carlesi2014,zivick_2015_grav,barreira_grav2015,massara_2015,Pisani_2015,Hamaus_2016,Hamaus_2017,pollina2016,voivodic_2016,Cautun_2018,Kreisch_2018,Schuster_2019,perico_2019, Pisani_2019, Verza_2019}. Voids span sizes of tens to hundreds of \hmpc. Their use for cosmology relies on averaged quantities such as their void size function or the stacked density profile (also known as the void-galaxy cross-correlation function)---needing large samples of voids. 

In cosmology, theoretical models for voids are mainly built and tested on halo catalogs from large dark matter (DM) N-body simulations (see e.g. \citet{hamaus_density}). In N-body simulations, halos are found using algorithms such as friends-of-friends \citep{gadget2001,gadget2009} and are then used as tracers to construct void catalogs. 
Ideally, model testing should be done on simulations able to mimic galaxy properties matching observations: mock galaxies would be ideal as tracers, since the ultimate goal is to apply those models to data \citep[e.g.][]{Hamaus_2016,Hamaus_2017,Hawken_2016}. 

In the best case, however, large N-body simulations are populated with Halo Occupation Distribution (HOD) techniques \citep{Zheng_2005} to replicate the properties of a particular galaxy population---this is still far from optimal \citep{Hadzhiyska_2020}, as it fails to include all the complicated effects impacting galaxy formation, and does not account for the impact of selection cuts performed on galaxies from observations. 
Typically, large simulations are unable to include the small scale evolution of stars and gas and fail to reproduce the observed population ratios of spiral and elliptical galaxies \citep{Vogelsberger_2014b}. Indeed modeling the baryonic component requires being able to mimic stars, gas, super-massive black holes, and their feedback. This is necessary to allow a robust prediction of galaxy properties (e.g. stellar content, morphology) on small scales and, consequently, to reliably reproduce galaxy populations over large scales.
Such effects are only accounted for in hydrodynamical simulations.

Current hydrodynamical simulations model galaxy populations and features in great detail (including, among others, effects such as primordial and metal-line gas cooling, star formation, stellar feedback, super-massive black hole formation, growth, and feedback), but the simulation of all these effects is computationally expensive and can only be performed for relatively small simulations sizes (e.g. \texttt{Illustris}, \cite{illustris:Genel-2014}) up to a few hundreds of \hmpc.
Consequently all recent works testing the extraction of cosmological information from voids rely on halo catalogues from DM simulations or HOD, implicitly assuming that the use of halos or HOD populated fields as tracers of the cosmic web---instead of galaxies---to measure void observables is robust, and that the impact of such assumption is small enough to be negligible (or, in any case, smaller than the error in the measurements of cosmological parameters).

As of today this assumption remains untested. With the increase in void numbers promised by upcoming surveys such as DESI \citep{DESI_2016}, Euclid \citep{Laureijs_2011}, SPHEREx \citep{Dore_2018} and WFIRST \citep{Spergel_2015}, errors in measurements of void related quantities will drop dramatically. The era of precision cosmology from voids steadily approaches: it becomes critical to test the impact of galaxy properties on void observables used for cosmology. Typically, when performing void finding in simulations we rely on halos above a certain mass. 

The mass of halos is often modeled as correlated with galaxy luminosity or galaxy star formation rate (SFR). Most N-body simulations will assume a linear relationship between the halo mass and the galaxy luminosity (or galaxy SFR) to build halo catalogs. Nevertheless it is well known that there is a considerable scatter in this relationship \citep{Greco2015,GuoWhite2011}. 
Very massive halos can correspond to low luminosity or low star formation rate galaxies. Voids are built using a tracer field; in observations the used tracers are galaxies, while in simulations we use halos. Void catalogs from simulations will then intrinsically be impacted by this scatter. To allow a proper understanding of application to data, such scatter needs to be quantified.  

In this work we build a tool to replicate the properties of the \texttt{Illustris} galaxy population on a larger simulation, the \texttt{MassiveNus} $\Lambda$CDM. 
We then construct void catalogs from \texttt{MassiveNus}, using as tracers both the DM halos and the reconstructed galaxies. This allows us to analyse void properties in both the original halo catalog and in the galaxy catalog obtained by replicating the properties of \texttt{Illustris} on the larger \texttt{MassiveNus} simulation. The comparison quantifies the impact on void observables of the scatter in the mass-to-luminosity (or SFR) relationship.  

The paper is shaped as follows: in Section \ref{sec:Simulation and Void Finder} we introduce the hydrodynamical and N-body simulations used for this work, as well as the void-finding algorithm. 
Section \ref{sec:Tracer Selection} presents the method used to replicate the properties of galaxies from \texttt{Illustris} in the large N-body \texttt{MassiveNus} simulation.
Section \ref{sec:Void Observables} describes the void observables on which we test the impact of the tracer used to build void catalogs, and presents our results.
We present our conclusions in Section \ref{sec:Conclusion}. 


\section{Simulations and Void Finder}
\label{sec:Simulation and Void Finder}

In this paper we rely on two simulations: the \texttt{Illustris} simulation provides us with a framework to reproduce galaxy properties relying on halos from the larger N-body simulation \texttt{MassiveNus}.

\subsection{Illustris}
The \texttt{Illustris} \citep{illustris:Vogelsberger-2014} project is a set of hydrodynamical particle simulations aiming to simultaneously be complex enough to model galaxy formation, large enough to model evolution on cosmic scales, and resolved enough so that small (as well as large) structures can be studied. The \texttt{Illustris} project grows out of previous work with the \texttt{Millennium} simulation and allows us to connect galactic scales to the large-scale structure of the universe\footnote{We note that a larger \texttt{IllustrisTNG} has been recently developed, with different sizes available  (50, 100 and 300 $Mpc$), extended mass range of galaxies and halos and other improvements \citep{illustris:Nelson-2015}---it would constitute a valid option to consider for further work.}. 

Cosmological parameters reflect a $\Lambda$CDM cosmology, with  $\Omega_{m} = 0.2726$, $\Omega_b = 0.0456$, and $\Omega_\Lambda = 0.7274$, $h=0.704$, $\sigma_8=0.809$, $n_s=0.963$; the parameters follow the Wilkinson Microwave Anisotropy Probe (WMAP)-9 measurements \citep{Hinshaw_2013}. Minimal changes of cosmological parameters are not expected to affect the results of the paper.  Initial conditions for the simulation are obtained with N-GenIC \citep{Springel_2005}\footnote{More details can be found here: \url{http://www.illustris-project.org/data/downloads/Illustris-1/}.}.
The simulations consists of a total of $1820^3$ dark matter particles and $1820^3$ gas tracer particles, each with a particle mass of $m_\mathrm{DM}=4.4 \times 10^6 \mathrm{M_\odot}/h$ in a 75 \hmpc-length cube ($\simeq 106.5 \mathrm{Mpc}$) \citep{\illustris} built using the moving mesh code AREPO \citep{Springel_2010}. The \texttt{Illustris} simulation includes modelling of subgrid physics, radiative gas cooling, star formation, galactic-scale winds from star formation feedback, super-massive black-hole formation, accretion and feedback. 
 The simulation provides halo catalogs built with friend-of-friends \citep{Davis_1985}. 
 Redshift snapshots span from $z=0.0$ to $z=127$.0. 
 For this work we train our models on the $z=0.0$ halos dataset, containing $7.7\times
 10^6$ halos.

\subsection{MassiveNus}
\texttt{MassiveNus}\footnote{\texttt{MassiveNuS} snapshots, halo catalogues, merger trees, and galaxy and CMB lensing convergence maps, are publicly available at \url{http://ColumbiaLensing.org}.} (Cosmological Massive Neutrino Simulations)\citep{massivenus:liu-2018}
are a set of 101 N-body simulations with size 512 \hmpc~ each. The simulations have a $\Lambda$CDM version, as well as boxes including the effect of massive neutrinos. Massive neutrinos modelling (for the boxes which we do not use in this work) is done by evolving neutrinos perturbatively and considering clustering with a non-linear cold DM potential. Other parameters are varied in the 101 versions: aside from the sum of neutrinos masses; the matter density $\Omega_m$ and the primordial curvature power spectrum $A_s$ are changed. Each box has $1024^3$ particles and the simulations span redshifts from 0 to 45. The publicly available simulated data includes simulation snapshots and halo catalogs that are relevant for our work. 

Initial conditions are obtained with an improved version of N-GenIC \citep{Springel_2005}, S-GenIC;  simulations are built with the public tree-Particle Mesh code Gadget-2 \citep{Springel_2005}. The halo catalogs are generated with the halo finder code Rockstar \citep{Behroozi_2013}, relying on a friends-of-friends algorithm that groups close particles and finds substructures within parent halos. The minimum halo mass in the simulations is roughly $10^{11} \mathrm{M}_\odot \cdot h^{-1}$ while the number of high mass halos ($> 10^{14}$) is slightly below model predictions due to the box size. 

Cosmological parameters for the massless neutrino $\Lambda$CDM simulation snapshot at $z=0$ used in this work are: $A_s$=2.1$\times 10^{-9}$, $\Omega_m$=0.3, $\Omega_{rad}=0$, $h$=0.7, $n_s$=0.97, $w$=$-1$, and $\Omega_b$=0.05. As for the \texttt{Illustris} simulation, we note that minimal changes of cosmological parameters are not expected to affect the results of the paper.
The volume is large enough to observe a considerable statistics of voids, while still resolving relatively low mass halos; this makes \texttt{MassiveNus} the most suitable simulation to test the impact of the halo mass-to-luminosity relationship scatter on void properties.

\subsection{Void Finder}
This project uses the \texttt{VIDE}\footnote{Publicly available at \url{https://bitbucket.org/cosmicvoids/vide_public}} void finder (\cite{VIDE}, embedding \texttt{ZOBOV} \cite{vide:Neyrinck-2008}), which operates in three steps: the tessellation of space, creation of a density field, and finally, a watershed transformation. The tessellation of space is performed as a Voronoi tessellation where each input particle, galaxy, or dark matter halo acts as a tracer of the large-scale structure.

The Voronoi tessellation \citep[see e.g.][]{Barr_2010} has been widely used in astrophysics  \citep[see e.g.][]{Ebeling_1993,weygaert2007voronoi,Weygaert_2009,Cappellari_2009,Soares_Santos_2010}. It associates to each tracer a cell (called a Voronoi cell) that contains all of the points that are closer to the tracer itself than any other tracer. Each cell is assigned a uniform density, inversely proportional to its volume. This creates a density field across all of space.
Cells corresponding to local density minima are merged with higher density cells surrounding them, until local density maxima are reached. The watershed transform effectively merges smaller basins into larger ones. The merged basins are the detected voids.

\texttt{VIDE} has been used in many recent applications for cosmology (examples include \cite{VIDE,Pisani_2015,Hamaus_2016,hamaus_rsd_2015,sutter2013aVoid_galaxy,pollina2017,Pollina_2018,Kreisch_2018,Schuster_2019,contarini2019, Verza_2019}), both to test theoretical models and to constrain parameters with data---it will likely be used in upcoming years to analyze data from future surveys and for model testing with future state-of-the-art mocks and simulations\footnote{Extensive tests of different techniques to find voids have been considered in the literature \citep[see e.g.][]{Colberg_2008,Cautun_2018}. Here we focus on the void finder \texttt{VIDE}: while we do expect that different void finders might be impacted differently, most recent applications using voids for cosmology rely on either \texttt{VIDE} or are based on \texttt{ZOBOV}---hence our results cover work from a broad range of groups.}. It is therefore important to check the impact of tracer selection properties on void finding.\footnote{An interesting application for further work would be to focus on analysing generalizations of the Voronoi smoothing technique (such as smoothers based on the empty space function F, see e.g. \citet{Baddeley_2015}), and, in particular the impact of such other techniques on the results of this work. Nevertheless we note that, if a different void finder were to be used, there would be no need to repeat the training, as the void finding procedure is performed after assigning galaxy properties to MassiveNus halos. If a different halo finder were to be used in the Illustris simulations, the mapping between galaxies and halos could change, but the scatter on the relationship between halos and galaxies should be preserved. In this case, re-running the training would ensure an enhanced robustness against changes due to the halo definition. Globally we do not expect changes in either the halo definition or the void definition to impact our overall results.}  


\section{Tracer Selection}
\label{sec:Tracer Selection}

This section describes the algorithm to reproduce the relationship between halos and galaxies modeled in \texttt{Illustris} on \texttt{MassiveNus} halos. \texttt{Illustris} properly takes into account the scatter in the relationship, thanks to the realistic implementation of galaxies' properties.
Rich data from \texttt{Illustris} allow us to bridge the computational gap between high-resolution, complex simulations and larger simulations with relatively lower resolution and complexity---namely, \texttt{MassiveNus}. We build a predictive model mapping a halo's mass to the properties of the galaxy within that halo. The model can then be utilized to probabilistically assign galaxy properties to a halo. 
It associates galaxy-like properties to pre-existent halos, by assigning galaxy features (luminosity and SFR), mapped from the halo mass, to existing halo positions.

In particular we focus on two galaxy properties that will impact selection when using real data: the luminosity of a galaxy and the star formation rate (SFR). Illustris allows us to link those properties to halo mass, reproducing the intrinsic scatter present in the relation.

We now describe how to build the model with \texttt{Illustris} and how to subsequently apply it to \texttt{MassiveNus}.
\subsection{Constructing the model}
Given the nature of the distribution of halo masses with luminosity and SFR, different groups of galaxies will have different probability distributions that accurately link halo masses to properties. To construct the model, we first split our sample of halo masses in even-width groups (with the same number of halos in each). Each group will therefore have an adequate number of data points. We use 5 halos per group, but tested different combinations.
Our goal is to find the optimal model for each group of galaxies.
To do so, independently for each group, we find the discrete random variable that describes the distribution of the desired property in that group.

To find the discrete random variable for each group, we consider the dark matter halo masses for each galaxy in the group as a random variable and fit a discrete distribution. We compute a histogram with 100 bins over the data within the group itself. In other words, we use a normalized histogram of the dark matter masses. As an example, if a bin with average dark matter mass $m$ contains half of the total halos in that group, then $P M F (m) = 0.5$. We repeat for each group to find the set of discrete random variables that best describes each case.

From the random variable associated with each group we can draw properties for each halo. Through this process, the two properties (luminosity and star formation rate) are independently assigned to the halos. 

The final model is then the set of random variables and the boundaries defining the groups.

\subsection{Applying the model}

\begin{figure}
  \centering
  \begin{tabular}[b]{@{}p{0.47\textwidth}@{}}
    \centering\includegraphics[width=1.05\linewidth]{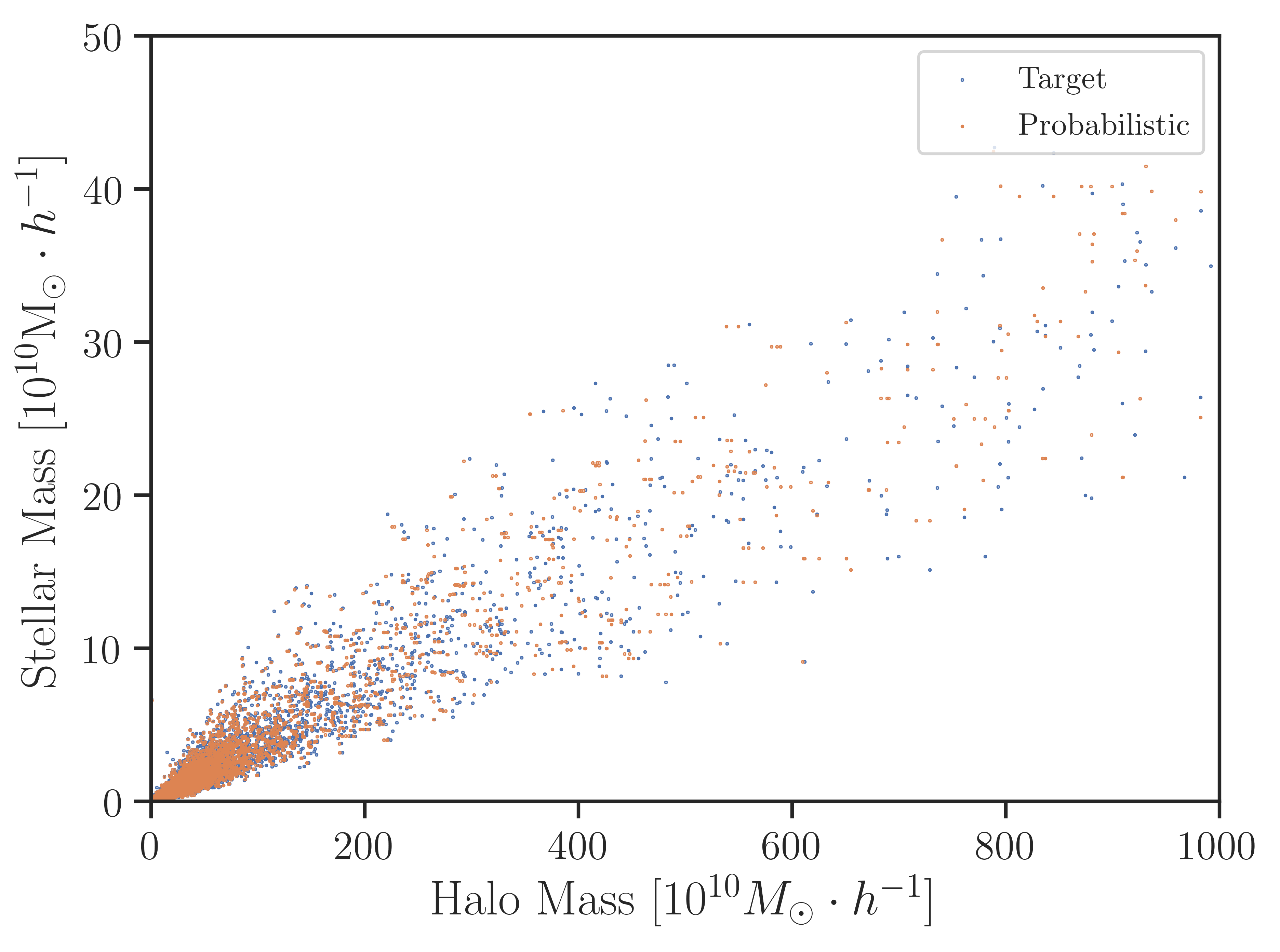} \\
  \end{tabular}%
    \hspace{-20pt}
  \begin{tabular}[b]{@{}p{0.47\textwidth}@{}}
    \centering\includegraphics[width=1.05\linewidth]{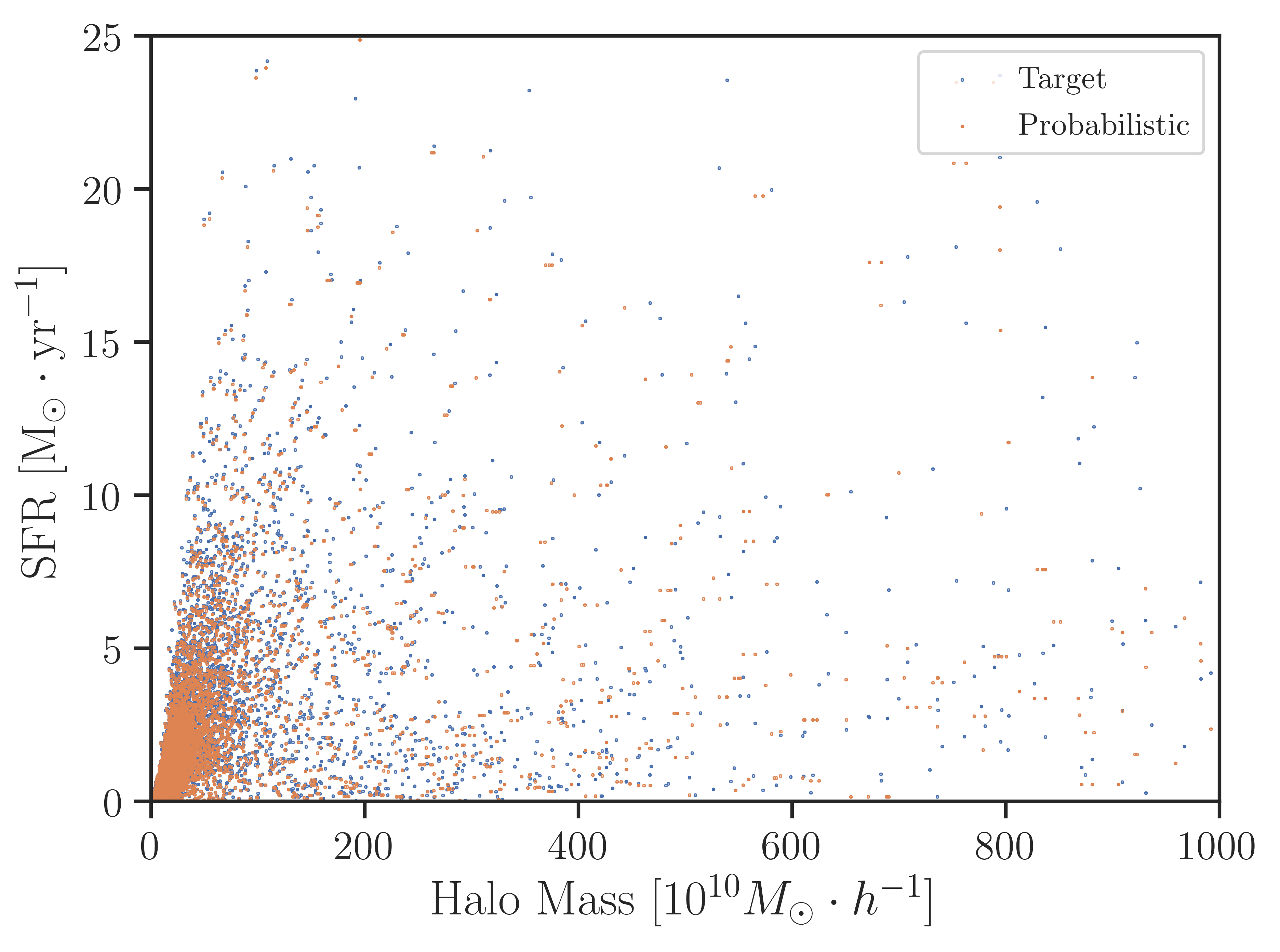} \\
  \end{tabular}
  
  \caption{Probabilistic fits for the luminosity (left) and star formation rate (SFR) (right) overlaid on top of the original data.}
   \label{fig:probabilisticfits}
\end{figure}

 After building the model, we first test it on \texttt{Illustris} data: Figure \ref{fig:probabilisticfits} shows the probabilistic fits for luminosity and star formation rates overlaid on top of original data from  \texttt{Illustris}. The probabilistic distribution is reproducing the target distribution for the whole range of halo masses in both the luminosity case and the SFR case, with a mean squared error of, respectively, $1.8 \times 10^{9}\,h^{-1}M_\odot$ and  $0.021 (M_\odot/ yr)^2$. 

Our model is subsequently applied to \texttt{MassiveNus} by fitting each halo within the original slices: the probabilistic approach allows to replicate both galaxies luminosities, or star formation rates on \texttt{MassiveNus} halos.

After assigning galaxy-like properties to halos, we now have three different sets of tracers from \texttt{MassiveNus} to perform void finding:

\begin{itemize}
    \item halo catalog: the original output from the \texttt{Rockstar} halo finder provided as part of \texttt{MassiveNus}
    \item luminosity ordered galaxy catalog: built with our probabilistic approach from \texttt{MassiveNus} halos
    \item SFR ordered galaxy catalog: built with our probabilistic approach from \texttt{MassiveNus} halos
\end{itemize}

We wish to replicate what is done in simulations and observations, to test and subsequently apply models. 

In simulations, a mass-cut is performed on the original set of halos. To replicate what is done in observations, we apply a cut on galaxy catalogs chosen in terms of descending luminosity or SFR, keeping a number of resultant galaxies equivalent to the number obtained from the simulation mass-cut. That is, if the simulation mass-cut results in $n$ halos, the galaxies with the $n$ highest luminosity or SFR, respectively, are taken. The cuts on galaxy-like catalogs mimic real data treatment. We note the importance of keeping the same number of tracers, as void number density can be sensitive to tracer number density, both in terms of minimum mean particle separation (impacting smaller voids to be detected on a tracer catalog) and of box length (impacting the statistic of large voids and, potentially, their observed maximum size).
We note that we perform two different mass cuts, and replicate tracer numbers in each case on the luminosity and SFR ordered tracer catalogs. The mass cuts are $M_\mathrm{min}=2.5 \times 10^{12}\,h^{-1}M_\odot$ and $M_\mathrm{min}=5 \times 10^{12}\,h^{-1}M_\odot$. 
 
We obtain six void catalogs from the tracers described in the items above (three kinds of tracers, considering two different mass cuts for each case). The comparison of void properties in the catalogs will allow us to estimate the impact of the scatter in the relationship between halo mass and luminosity or SFR selected tracer catalogs. 
 
In the next section we describe the two main void observables we focus our attention on, the void size function and the void density profile, and we present our results.

\vspace{30pt}
\section{Results: impact on voids}
\label{sec:Void Observables}

\subsection{Void size function}

The void size function---the number of voids as a function of their radius, also known as void abundance---is a sensitive probe of cosmology. In particular it is sensitive to the properties of dark energy, modified gravity and neutrinos \citep[e.g.][]{Pisani_2015,zivick_2015_grav,massara_2015,Sahlen_2015,Sahlen_2018,Kreisch_2018, Schuster_2019, Pisani_2019, Verza_2019}. 

Aside from being a relevant cosmological observable, the void size function provides an estimate of the number of voids to be observed by an hypothetical survey. Since the number of voids will impact error-bars on measurements of the density profiles or other void features sensitive to cosmology, it is important that the void size function is modeled correctly. Recent developments have established a robust framework to predict void abundance based on an improved version of the popular Sheth and Van de Weygaert model (an excursion set model, \cite{Sheth_2003}): the volume conserving model (Vdn, \cite{jennings}). Such a model has proven successful in predicting void numbers from dark matter simulations, using both dark matter and halos (see recent results in \cite{Ronconi_2019, contarini2019, Verza_2019}), and accounting for halo bias in standard $\Lambda$CDM and different dark energy models\footnote{Interestingly, the match with theoretical predictions is done with the same void finder \texttt{VIDE} that we use in our work.}.

The impact on void abundance of the scatter in the relationship between halo mass and luminosity or SFR selected tracer catalogs has yet to be estimated---a relevant point to consider as the void size function is becoming an established tool to use with upcoming data. 

\begin{figure}
  \centering

  \begin{tabular}[b]{@{}p{0.47\textwidth}@{}}
    \centering\includegraphics[width=1.05\linewidth]{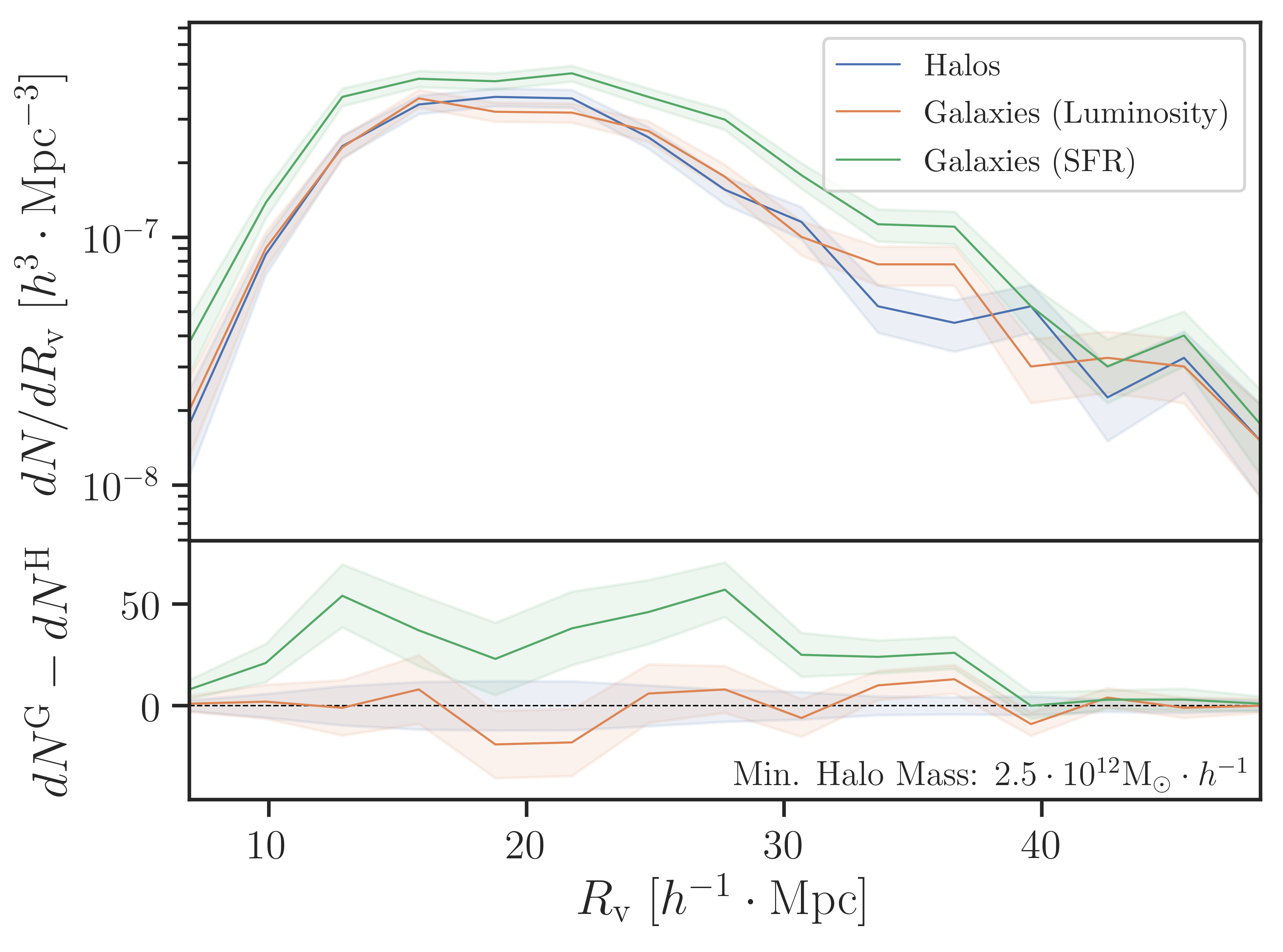} \\
  \end{tabular}%
  \hspace{-20pt}
  \begin{tabular}[b]{@{}p{0.47\textwidth}@{}}
    \centering\includegraphics[width=1.05\linewidth]{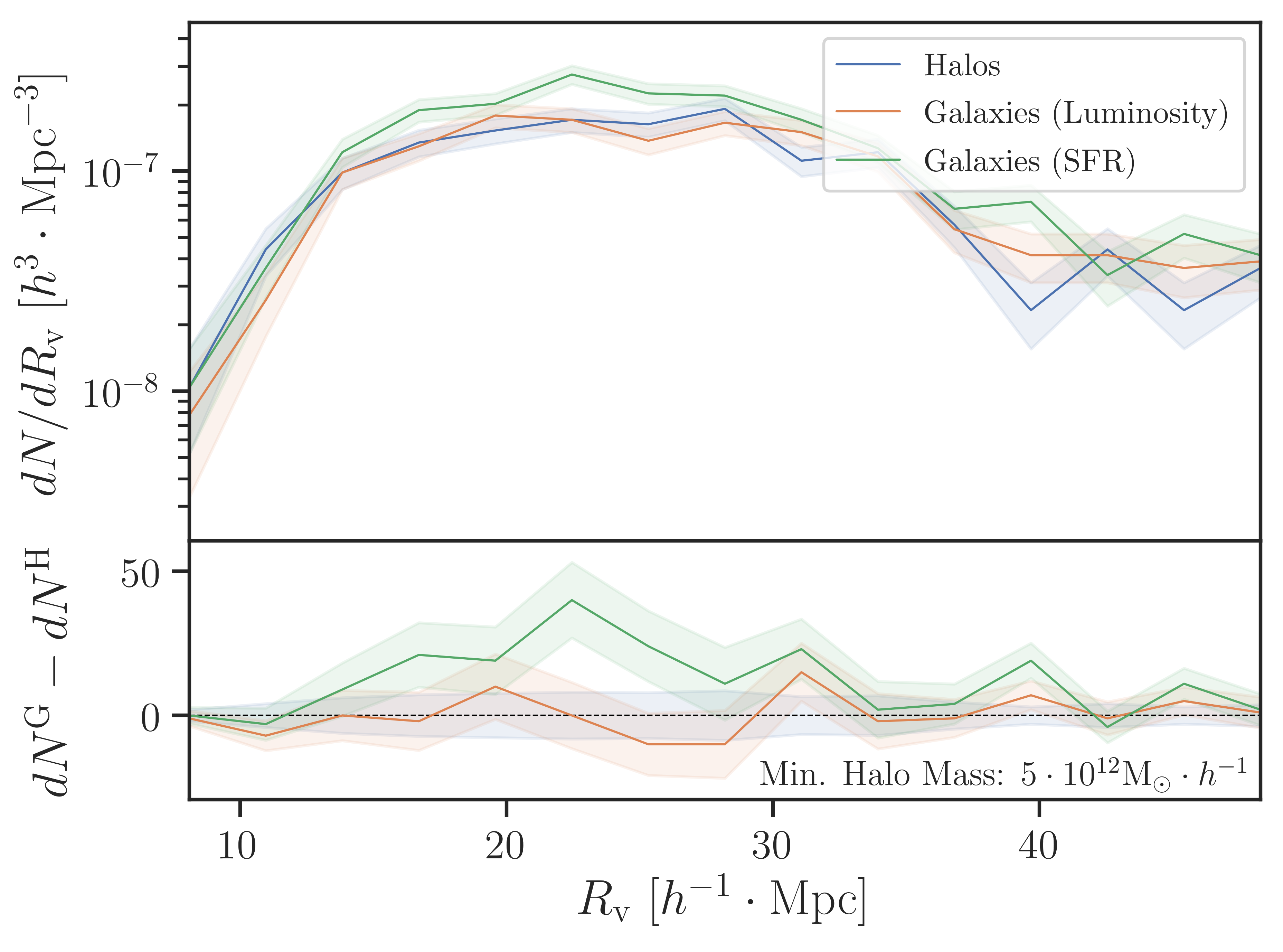} \\
  \end{tabular}
  
 \vspace{-10pt}
  \caption{Void abundances for the $2.5 \times 10^{12} h^{-1}M_\odot$ set (left) and the $5 \times 10^{12} h^{-1}M_\odot$ set (right). Bottom panels show the relative differences between halo-traced and galaxy-traced cases. The shaded region for each abundance is the Poisson error.}
  \label{fig:results:void-abundances}
\end{figure}

We run the void finder on all the tracer catalogs described in Section \ref{sec:Tracer Selection} and compare the results. 
We show in Figure~\ref{fig:results:void-abundances} the comparison of the void size function. Error bars are obtained considering Poisson error.
Within error-bars, voids obtained from the halo distribution have the same void size function as voids selected from galaxies using a luminosity cut. Consistency remains for the two different mass cuts used, showing that testing theory and models on voids from halo catalogs can be used as a reliable proxy of data application, where galaxies are selected through their luminosity. 
We conclude that \textit{void abundances are robust against the scatter present in the relationship between the dark matter halo mass and the luminosity of galaxies}.

Given the volume of the \texttt{MassiveNus} simulation, the tracer number densities for the $M_\mathrm{min}=2.5 \times 10^{12}\,h^{-1}M_\odot$ tracer catalog and the $M_\mathrm{min}=5 \times 10^{12}\,h^{-1}M_\odot$ correspond, respectively, to a tracer number density of $\simeq 1.9 \times 10^{-3}$ $h^{3}\mathrm{Mpc}^{-3}$ and $\simeq 9.7 \times 10^{-4}$ $h^{3}\mathrm{Mpc}^{-3}$. These number densities roughly span the ranges of tracer densities to be observed by upcoming surveys (including, for example, DESI, PFS, Euclid or WFIRST---for some of the redshift bins, since the latter is expected to exceed these number densities for the most populated bins). This work thus mimics the level of precision that can be expected, for a given redshift bin, with future surveys, and shows that void abundance measurements will be robust against the scatter present in the halo mass-to-galaxy luminosity relationship. 

On the other hand, the selection of galaxies using SFR provides a void size function with higher values compared to voids from halos. While the global trend is similar for void abundance from halos and void abundance from SFR selected galaxies, the overall value is different, and is due to an overall higher void number in the SFR galaxies case.  

The larger number of voids in the SFR catalogs can be attributed to a bias in how tracers are selected for this type of threshold (see Section \ref{sec:Tracer Bias}). The galaxies in this catalog are chosen by taking the $n$ galaxies with the highest star formation rates. 

Galaxies within clusters have low star formation rates, compared to those that are not in clusters, because they have less gas available from which to form stars. Thus, the galaxies with the highest star formation rates, which are precisely the ones chosen for these catalogs, will be those outside of clusters; the SFR-cut tracer catalogs are biased towards galaxies that are distributed more sparsely. This is confirmed, for example, by the lower value of the two-point correlation function for the probabilistic SFR galaxy catalogs with respect to the two-point correlation function of halos (which we do not show).

The effect of selecting a more sparsely distributed set of tracers will result in splitting up larger voids by adding galaxies inside them, and consequently increase the overall void statistics. Since the SFR-cut tracer catalogs are biased towards galaxies that are distributed more sparsely, the splitting of large voids can be linked to the mass of the host halos. Galaxies within clusters have low star formation rates, therefore SFR ordered galaxies (with higher SFR) are associated to halos with lower masses that have lower bias.

We note that for the $M_\mathrm{min}=5 \times 10^{12}\,h^{-1}M_\odot$ tracer catalog, the increase in void abundance is lower, confirming the fact that it is due to a bias tracer selection effect. Indeed with the $M_\mathrm{min}=5 \times 10^{12}\,h^{-1}M_\odot$ tracer catalog, we are selecting more highly biased tracers, making the splitting of voids harder, and hence resulting in a lower difference in void abundance.
Since current void size function models take into account the bias of tracers to build the theoretical prediction, we expect this shift to be correctly modeled---provided that the tracer bias can be either measured or theoretically predicted. 

\subsection{Density Profiles}

We now estimate the impact of the scatter in the relationship between halo mass to luminosity and SFR of tracers for void density profiles. 
Density profiles of voids are mathematically equivalent to the cross-correlation between void centers and tracers (galaxies, halos, dark matter particles) \citep[e.g.][]{Cai_2016, Hamaus_2017,Massara_2018},
representing the variation of density contrast as a function of the distance from the void center. 
The void-tracer cross-correlation function is then under-dense in the center, the density increases towards void edges, up to the over-dense shell that surrounds the void \citep{Sheth_2003, paz_rsd_2013,Hamaus_2014,Pisani_2014}. 

\begin{figure}
  \centering

  
  \begin{tabular}[b]{@{}p{0.46\textwidth}@{}}
    \centering\includegraphics[width=1.05\linewidth]{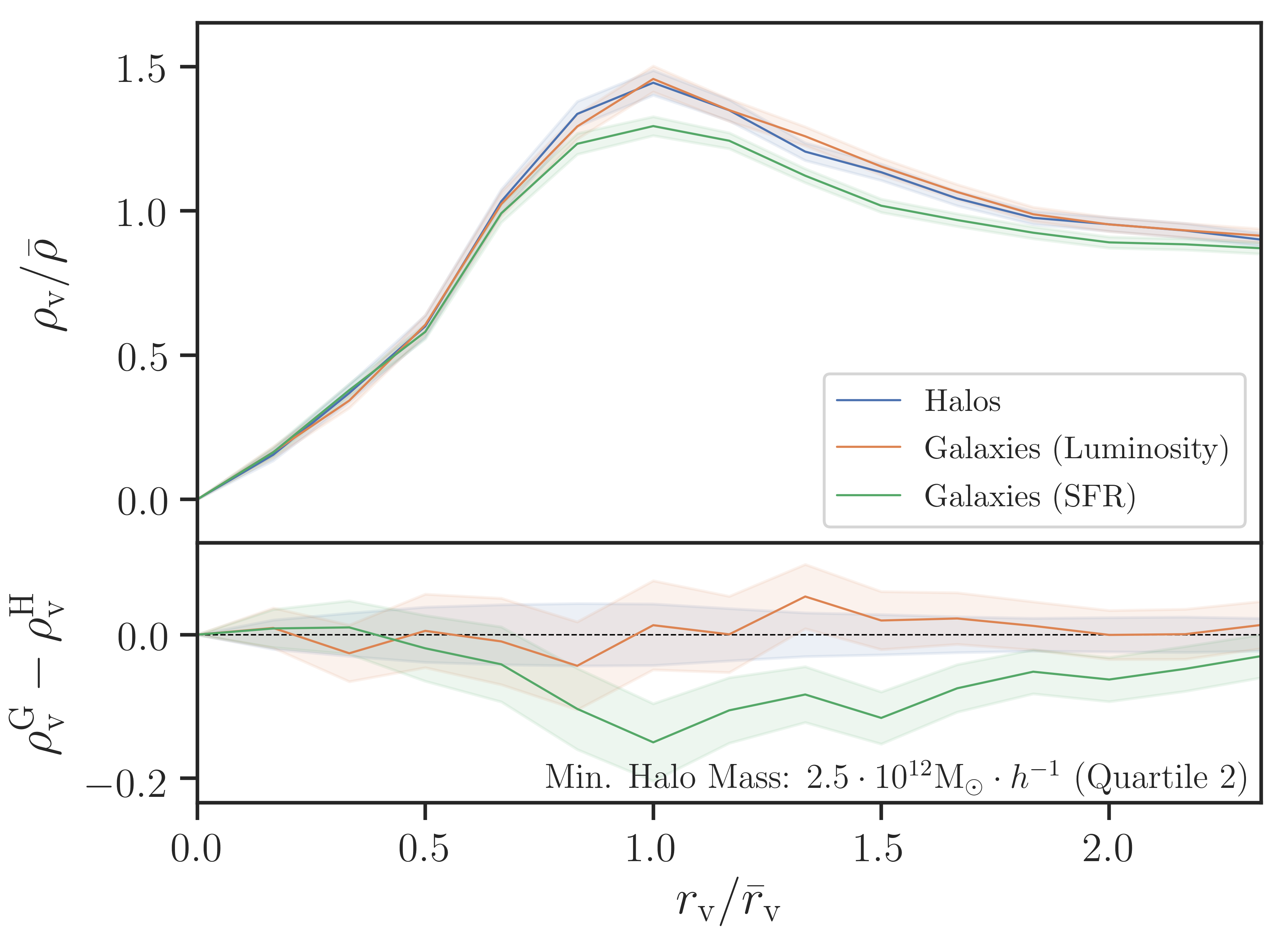} 
    \includegraphics[width=1.05\linewidth]{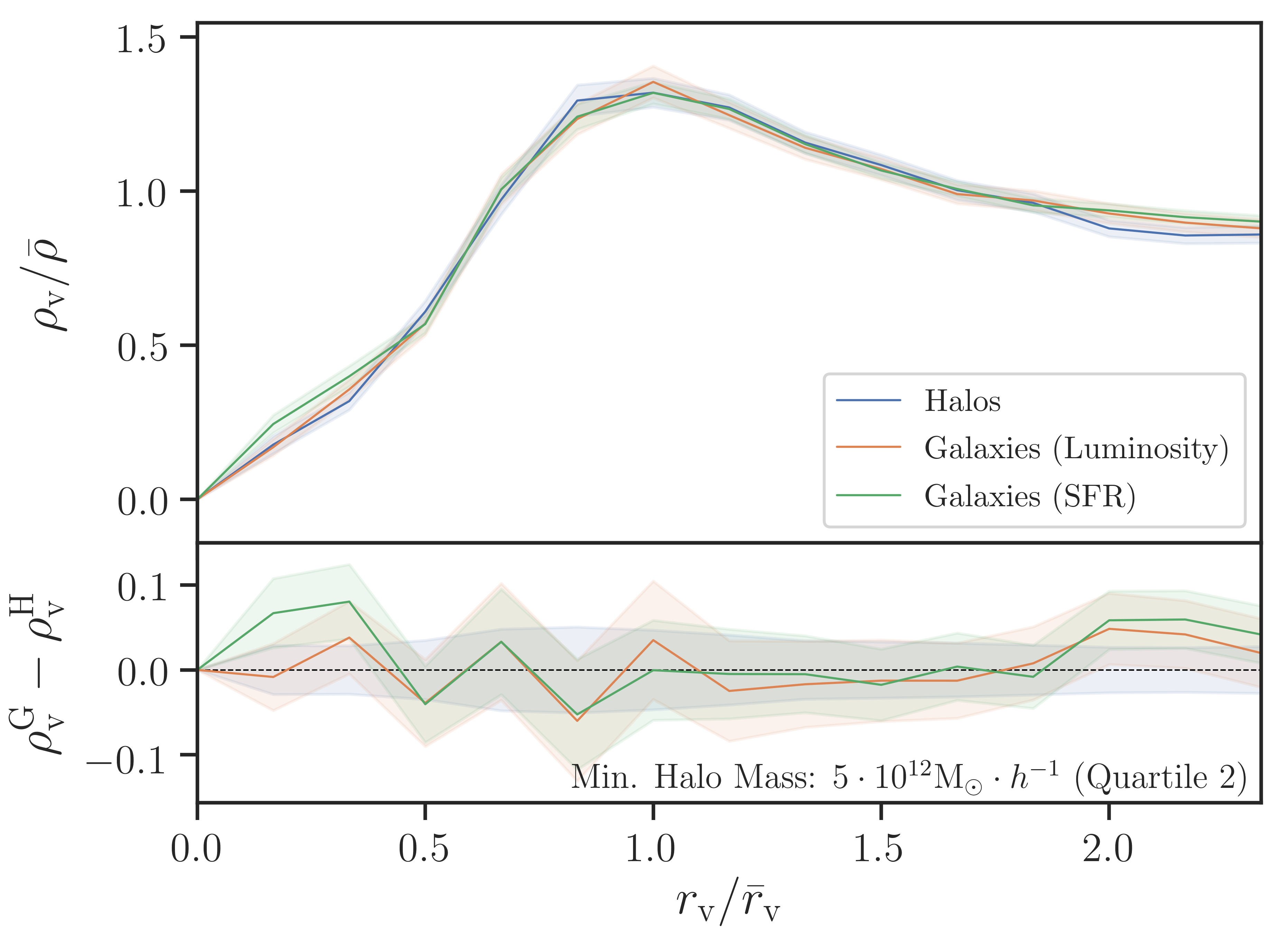}\\
  \end{tabular}%
      \hspace{-15pt}
  \begin{tabular}[b]{@{}p{0.46\textwidth}@{}}
    \centering\includegraphics[width=1.05\linewidth]{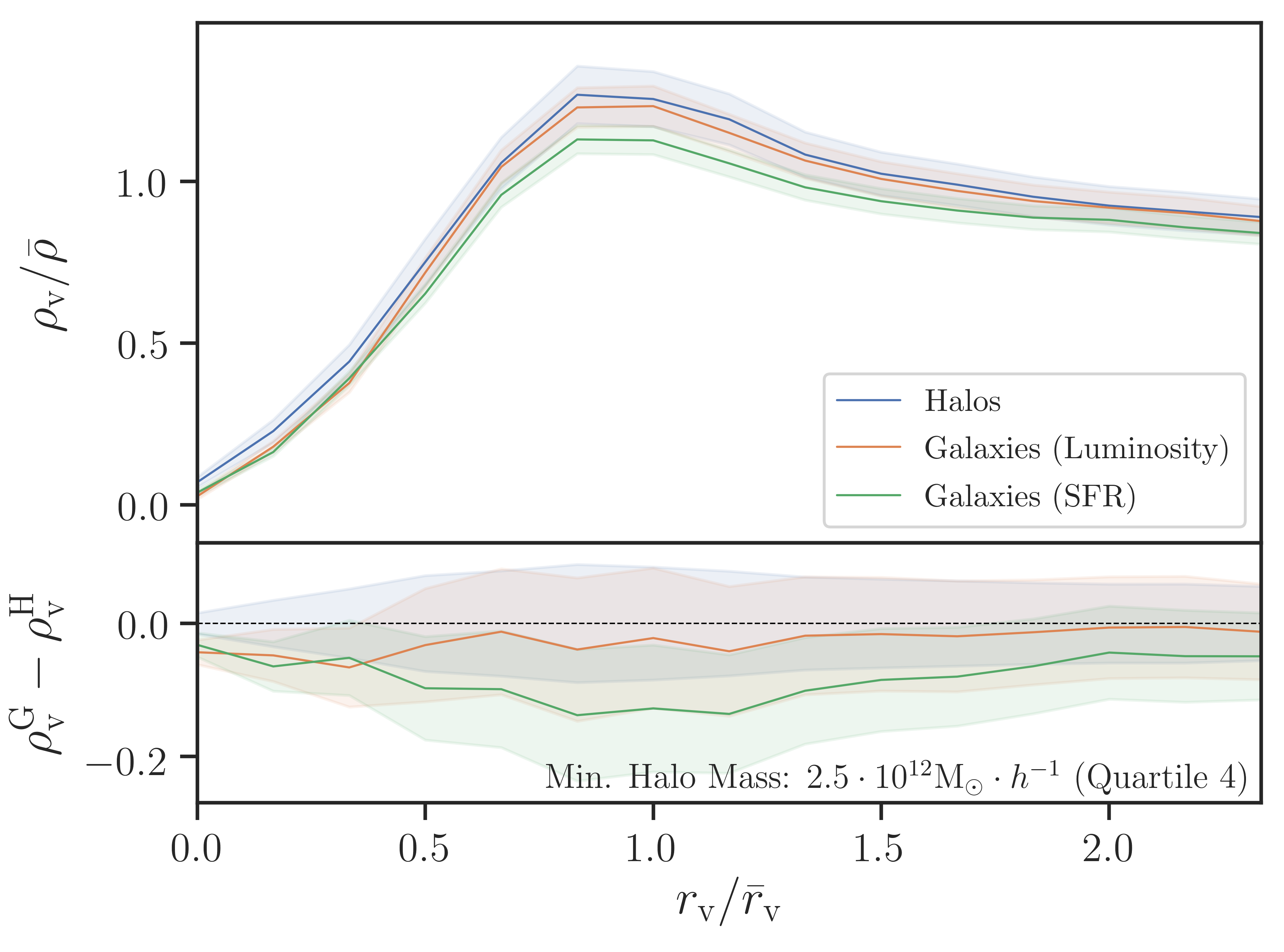} 
    \includegraphics[width=1.05\linewidth]{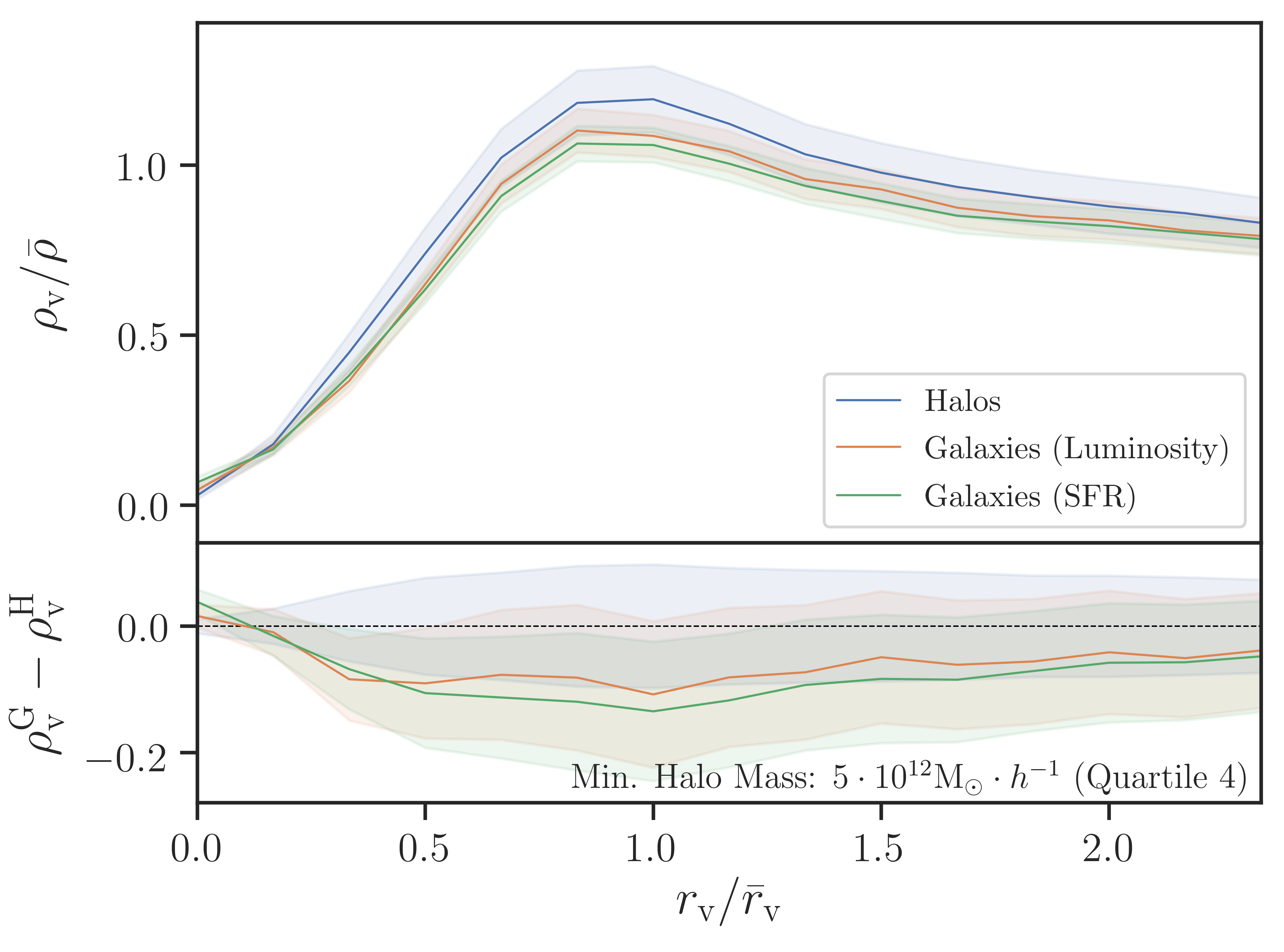}\\
  \end{tabular}

  \caption{Density profiles, of the second (left) and fourth (right) quartiles of void radii, for the $2.5 \times 10^{12} M_\odot\cdot h^{-1}$ set (top panels) and $5 \times 10^{12} h^{-1}M_\odot$ set (bottom panels). Bottom sub-plots for each panel show the relative differences between halo-traced and galaxy-traced cases. The shaded region for each profile is the error as the normalized standard deviation.}
 \label{fig:results:void-profiles}
\end{figure}

Void density profiles are powerful tools to constrain cosmology: the shape of void stacks is used as a standard sphere for the Alcock-Paczy\'nski test \citep{Lavaux_2012, sutter2014d_APtest, Hamaus_2016}, or to analyse the pattern of redshift-space distortions around voids and constrain the growth rate of structure \citep{Cai_2016,Hamaus_2017}. The measure of the void-galaxy cross-correlation function is, as of today, the only void observable that provided constraints on cosmological parameters from data \citep[e.g.][]{Hamaus_2016, Hawken_2016,Hamaus_2017, achitouv_2017, Achitouv_2019}.

We measure the density profiles from the different void catalogs described in Section \ref{sec:Tracer Selection}.
Figure \ref{fig:results:void-profiles} shows the results for different radius bins.\footnote{Note that we split in quartiles, for the $M_\mathrm{min}=2.5 \times 10^{12}\,h^{-1}M_\odot$ case, quartile 2 includes voids from roughly 16-21 \hmpc~and quartile 4 includes voids above 26 \hmpc~; for the $M_\mathrm{min}=5 \times 10^{12}\,h^{-1}M_\odot$ case, quartile 2 includes voids from roughly 20-26 \hmpc~and quartile 4 includes voids above 32 \hmpc.} Error bars are obtained considering Poisson error.
We note that profiles from different tracers are mostly consistent within error-bars, with some exception for the $M_\mathrm{min}=2.5 \times 10^{12}\,h^{-1}M_\odot$ case, where the SFR-galaxy based catalog results in a lower wall. This is consistent with the interpretation provided for the void size function: with a lower mass cut on halos, the SFR-galaxy catalog will be biased towards galaxies residing closer to void centers, hence selecting a different population of voids, defined by less biased tracers and hence with lower walls. For this reason the SFR void density profile has a lower ridge. 

This difference disappears when using the $M_\mathrm{min}=5 \times 10^{12}\,h^{-1}M_\odot$: in this case profiles from the different cases overlap within error-bars, as expected. Two reasons can explain the better match in this case. First, the $M_\mathrm{min}=5 \times 10^{12}\,h^{-1}M_\odot$ case has less voids, and hence larger error-bars, than  the $M_\mathrm{min}=2.5 \times 10^{12}\,h^{-1}M_\odot$ case, mitigating the differences between the halos and galaxy cases. Second, since the $M_\mathrm{min}=5 \times 10^{12}\,h^{-1}M_\odot$ case selects more strongly biased galaxies as a starting tracer catalog (than the $M_\mathrm{min}=2.5 \times 10^{12}\,h^{-1}M_\odot$ case) the impact of the SFR selection---which selects less biased objects among the available ones---is lower.

We conclude that \textit{void density profiles are robust against the scatter present in the relationship between the dark matter halo mass and the luminosity of galaxies}. For SFR selected galaxies, void density profiles will also be robust for upcoming surveys with tracer number densities lower than $\simeq 1 \times 10^{-3}$ $h^{3}\mathrm{Mpc}^{-3}$ (corresponding to the $M_\mathrm{min}=5 \times 10^{12}\,h^{-1}M_\odot$ case), while void profiles from denser surveys will only be robust if galaxies are luminosity or flux-selected. 

In the next section we wish to verify that bias is indeed different for galaxies selected with respect to luminosity or SFR, to confirm our interpretation of void size function and density profiles differences.

\subsection{Tracer Bias}
\label{sec:Tracer Bias}
\begin{figure}
  \centering
  
  
  \begin{tabular}[b]{@{}p{0.46\textwidth}@{}}
   \centering\includegraphics[width=1.05\linewidth]{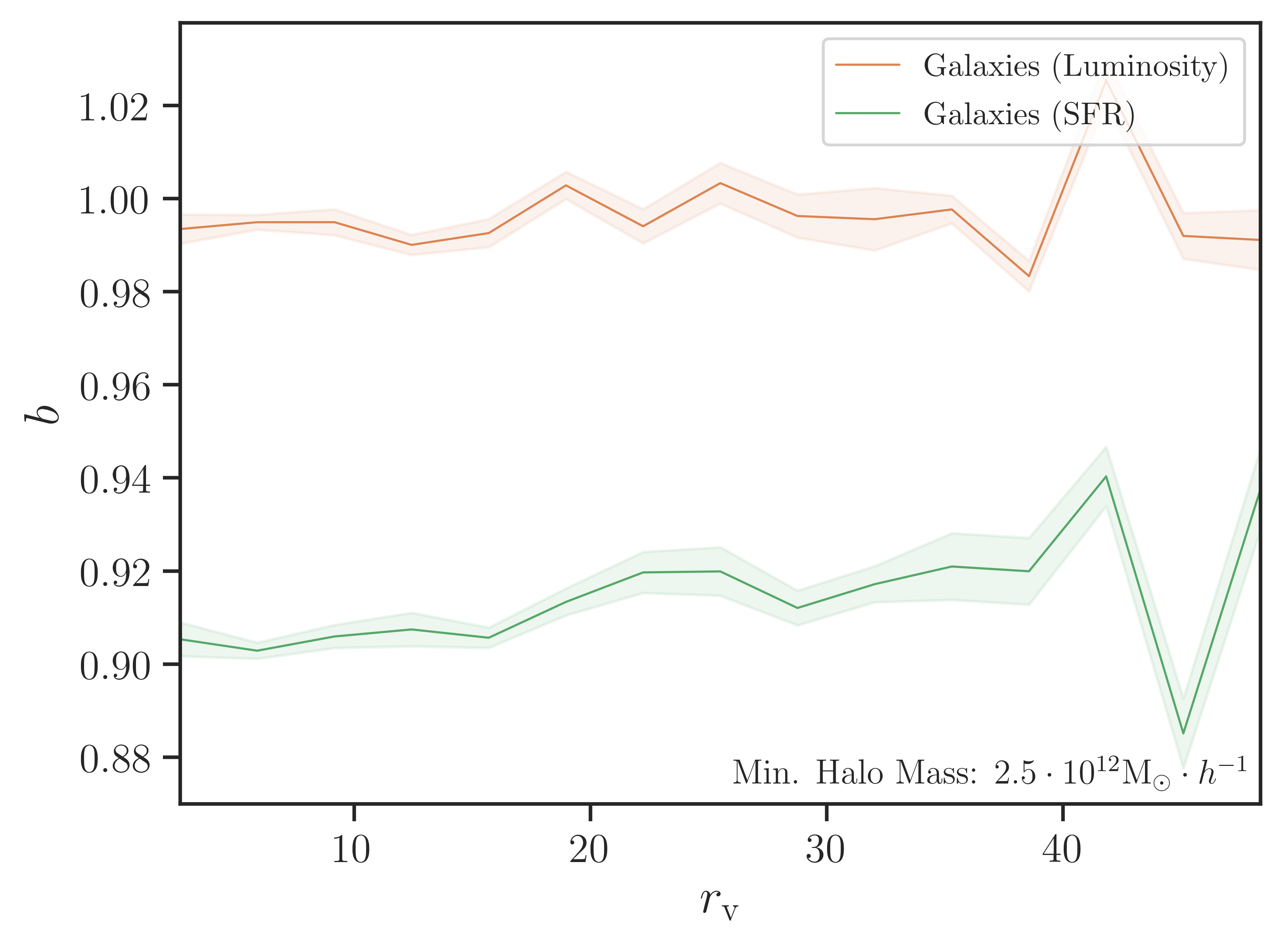} \\
  \end{tabular}%
        \hspace{-15pt}
  \begin{tabular}[b]{@{}p{0.46\textwidth}@{}}
    \centering\includegraphics[width=1.05\linewidth]{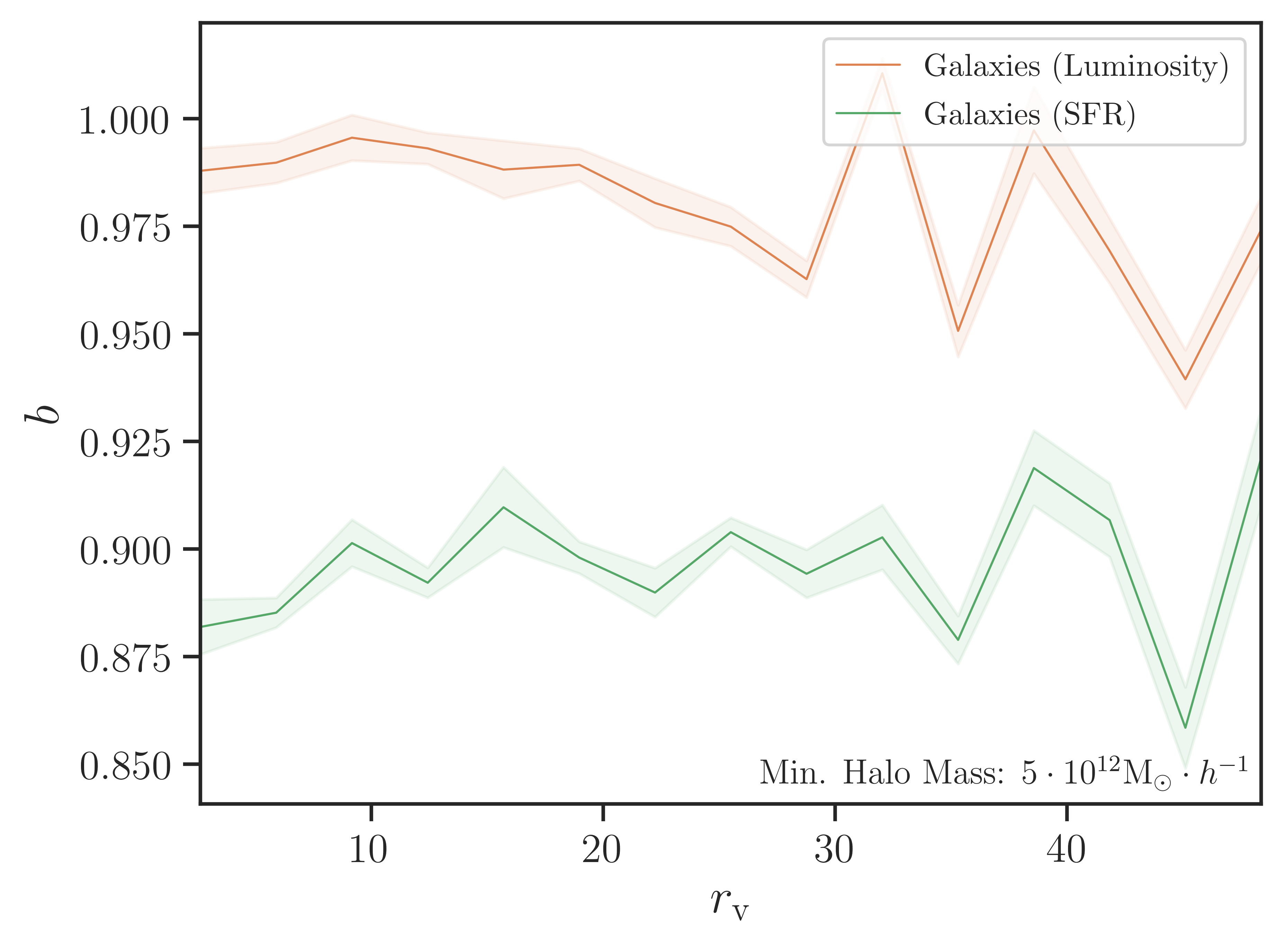} \\
  \end{tabular}
  \caption{Relative tracer bias for the $2.5 \times 10^{12} h^{-1}M_\odot$ set (left) and $5 \times 10^{12} h^{-1}M_\odot$ set (right).}
\label{fig:results:tracer-bias}
\end{figure}

We calculate the relative bias of the galaxies selected by luminosity and by SFR with respect to halos (error bars are obtained through jackknife resampling). As expected Figure \ref{fig:results:tracer-bias} shows that SFR selected galaxies have a lower bias than the luminous-mass selected galaxies, hence justifying the observed higher abundance and the density profiles with lower walls. 

\vspace{25pt}
\section{Conclusion and Discussion}\label{sec:Conclusion}

In this work we have tested the assumption that the void size function and the void-galaxy cross-correlation function are robust against the scatter in the halo mass-to-luminosity, or halo mass-to-SFR relationship for tracers used to build void catalogs. 
Mimicking on the large N-body simulation \texttt{MassiveNus} the scatter observed in the \texttt{Illustris} simulation, we find that both the abundance and the void density profiles remain consistent within error-bars for galaxies selected through luminosity. This result sets the ground to use void properties in upcoming data and robustly apply models developed with halos from N-body simulations.  
We note that using larger simulation volumes would reduce error-bars, but also likely lead to a lower number density of tracers due to the decrease in resolution; this would mimic a harsher luminosity cut, and hence a stronger bias: the effects of the SFR selection should be increasingly mitigated. 

Future work applications would involve the use of machine learning techniques such as \cite{Siyu_2018} to improve the population of large N-body simulations with galaxies, as well as testing our results on larger simulations---provided that a sufficiently low halo mass is reached even with simulations spanning larger volume (that is, maintaining a high resolution). While the void cross-correlation function is already currently used to extract competitive cosmological constraints from voids (e.g. relying on SDSS data, VIPERS data), further relevant steps that we leave for future work (such as the impact of survey mask, and of peculiar velocities) remain to be investigated and thoroughly modeled to reliably use the void size function.

\section*{Acknowledgements}
The authors are grateful to Elena Massara, M\'elanie Habouzit, Nico Hamaus, Jia Liu and Shy Genel for useful discussions. The authors thank the anonymous referees for their helpful comments. AP is supported by NASA grant 15-WFIRST15-0008 to the Nancy Grace Roman Space Telescope Science Investigation Team ``Cosmology with the High Latitude Survey''.
For using \texttt{MassiveNuS} in our work, we thank the Columbia Lensing group (\url{http://columbialensing.org}) for making their suite of simulations available.

\bibliography{Bibliography} 




\end{document}